\documentclass[aps,prl,preprint,tightenlines,superscriptaddress,showpacs,byrevtex]{revtex4}

\usepackage{graphicx}

\usepackage{graphicx}
\usepackage{dcolumn}
\usepackage{bm}

\def\to {\rightarrow}
\def\br           {{\cal{B}}}
\def\bdthrpi      {        {B^0 \to D^{*-}\,(3\pi)^+}}
\def\bdzethrpi    {        {B^+ \to \overline{D}^{*0}\,(3\pi)^+}}
\def\bdfourpi     {        {B^+ \to  D^{*-} (4\pi)^{++}}}
\def\bdzefourpi   {        {B^0 \to \overline{D}^{*0}\:(4\pi)^0}}
\def\bdfivepi     {        {B^0 \to D^{*-}\,(5\pi)^+}}
\def\bdzefivepi   {        {B^+ \to \overline{D}^{*0}\:(5\pi)^+}}

\begin{document}

\title{

\Large \bf Observation of $ \bdfivepi, \: \bdfourpi$ and $\bdzefivepi$}

\affiliation{Budker Institute of Nuclear Physics, Novosibirsk}
\affiliation{Chonnam National University, Kwangju}
\affiliation{University of Cincinnati, Cincinnati, Ohio 45221}
\affiliation{Deutsches Elektronen--Synchrotron, Hamburg}
\affiliation{University of Frankfurt, Frankfurt}
\affiliation{Gyeongsang National University, Chinju}
\affiliation{University of Hawaii, Honolulu, Hawaii 96822}
\affiliation{High Energy Accelerator Research Organization (KEK), Tsukuba}
\affiliation{Hiroshima Institute of Technology, Hiroshima}
\affiliation{Institute of High Energy Physics, Vienna}
\affiliation{Institute for Theoretical and Experimental Physics, Moscow}
\affiliation{J. Stefan Institute, Ljubljana}
\affiliation{Kanagawa University, Yokohama}
\affiliation{Korea University, Seoul}
\affiliation{Kyungpook National University, Taegu}
\affiliation{Swiss Federal Institute of Technology of Lausanne, EPFL, Lausanne}
\affiliation{University of Ljubljana, Ljubljana}
\affiliation{Nagoya University, Nagoya}
\affiliation{Nara Women's University, Nara}
\affiliation{National Central University, Chung-li}
\affiliation{National United University, Miao Li}
\affiliation{Department of Physics, National Taiwan University, Taipei}
\affiliation{H. Niewodniczanski Institute of Nuclear Physics, Krakow}
\affiliation{Nihon Dental College, Niigata}
\affiliation{Niigata University, Niigata}
\affiliation{Osaka City University, Osaka}
\affiliation{Osaka University, Osaka}
\affiliation{Panjab University, Chandigarh}
\affiliation{Peking University, Beijing}
\affiliation{Princeton University, Princeton, New Jersey 08545}
\affiliation{Saga University, Saga}
\affiliation{University of Science and Technology of China, Hefei}
\affiliation{Sungkyunkwan University, Suwon}
\affiliation{University of Sydney, Sydney NSW}
\affiliation{Tata Institute of Fundamental Research, Bombay}
\affiliation{Toho University, Funabashi}
\affiliation{Tohoku Gakuin University, Tagajo}
\affiliation{Tohoku University, Sendai}
\affiliation{Department of Physics, University of Tokyo, Tokyo}
\affiliation{Tokyo Institute of Technology, Tokyo}
\affiliation{Tokyo Metropolitan University, Tokyo}
\affiliation{Tokyo University of Agriculture and Technology, Tokyo}
\affiliation{University of Tsukuba, Tsukuba}
\affiliation{Virginia Polytechnic Institute and State University, Blacksburg, Virginia 24061}
\affiliation{Yonsei University, Seoul}

  \author{G.~Majumder}\affiliation{Tata Institute of Fundamental Research, Bombay} 
  \author{K.~Abe}\affiliation{High Energy Accelerator Research Organization (KEK), Tsukuba} 
  \author{K.~Abe}\affiliation{Tohoku Gakuin University, Tagajo} 
  \author{I.~Adachi}\affiliation{High Energy Accelerator Research Organization (KEK), Tsukuba} 
  \author{H.~Aihara}\affiliation{Department of Physics, University of Tokyo, Tokyo} 
  \author{M.~Akatsu}\affiliation{Nagoya University, Nagoya} 
  \author{Y.~Asano}\affiliation{University of Tsukuba, Tsukuba} 
  \author{T.~Aziz}\affiliation{Tata Institute of Fundamental Research, Bombay} 
  \author{S.~Bahinipati}\affiliation{University of Cincinnati, Cincinnati, Ohio 45221} 
  \author{A.~M.~Bakich}\affiliation{University of Sydney, Sydney NSW} 
% ...

\collaboration{The Belle Collaboration}

\begin{abstract}
          We report the first observation of a number of decay modes
          of the $B$ meson, namely $\bdfivepi$, $\bdfourpi$ and 
          $\bdzefivepi$, where $(n\pi)$ implies the combination of $n$
          charged pions. The analysis is based on a 140 fb$^{-1}$ data
          sample collected at the $\Upsilon (4S)$ resonance with the
          Belle detector at KEKB. We measure $\br (\bdfivepi)$ =
          $(4.72 \pm 0.59 \pm 0.71) \times 10^{-3}$, $\br (\bdfourpi )$ = 
          $(2.56 \pm 0.26 \pm 0.33) \times 10^{-3}$ and $\br (\bdzefivepi )$
          = $(5.67 \pm 0.91 \pm 0.85) \times 10^{-3}$. We also provide
          improved measurements of the branching fractions for the decay modes 
          $\bdthrpi$, $\bdzethrpi$ and $\bdzefourpi$.

\end{abstract}
\maketitle

A large fraction ($\sim 35\%$) of $B$ meson decays
is due to decay modes which are still unknown. While a
sizable fraction of $B$'s decay into semileptonic final states
($\sim 24\%$)~\cite{pdg2002}, decays into hadronic final states are dominant.
At present, however, only half of these final states
correspond to measured exclusive hadronic decays.

On average, $B$ mesons decay into a large number
of particles: the mean charged particle multiplicity in
hadronic $B$ decay is measured to be $5.8 \pm 0.1$~\cite{cleochg}.
Decay products of $D$ mesons make a significant contribution,
but $B$ decays to charmed states with a large number of
accompanying pions, $B \to \overline{D}^{(*)} (n\pi)$,
where the $\pi$ are charged pions and $n=3,4$, are also known~\cite{cleorhoa1}.
The invariant mass distribution of the multi-pion system,
and the resonant decomposition for such decays,
are important for the study of factorization~\cite{llw}.

In this paper, we present a study of inclusive
$B \to \overline{D}^{*0(-)} (n\pi)$ final states,
where $n = 3, 4$ and $5$, and 
measure branching fractions for six decay modes. 
Inclusion of charge conjugate modes is implied throughout this paper.

The analysis is based on a $140\,\mathrm{fb}^{-1}$ data sample
at the $\Upsilon (4S)$ resonance (10.58 GeV) 
and a 16 fb$^{-1}$ data sample 60 MeV below the $\Upsilon (4S)$ peak 
(referred to as {\it off-resonance} data), collected by the Belle detector 
\cite{belledet} at 
the asymmetric $e^+e^-$ collider KEKB \cite{kekb}. The data 
sample contains  152 million $B\overline{B}$ events. 

The Belle detector is a general purpose magnetic spectrometer 
with a 1.5 Tesla magnetic field provided by a superconducting solenoid.
Charged particles are measured using a 50 layer Central Drift Chamber 
(CDC) and a 3 layer double sided Silicon Vertex Detector (SVD). Photons 
are detected in an electromagnetic calorimeter (ECL) consisting of
8736 CsI(Tl) crystals.
Exploiting the information acquired from an array of 128 time-of-flight
counters (TOF), an array of 1188 silica aerogel 
\v{C}erenkov threshold counters (ACC) and $dE/dx$-measurements in CDC we
derive particle identification likelihoods 
${\mathcal{L}}_{\pi/K}$.
A kaon candidate is identified by a requirement on the likelihood ratio
${\mathcal{L}}_K/({\mathcal{L}}_K+{\mathcal{L}}_\pi)$ such that the average
 kaon identification efficiency is $\sim$ 90\% and the pion fake 
rate is $\sim$ 9\%. Similarly charged pions are selected with an
efficiency of $\sim$ 91\% and the kaon fake rate is $\sim$ 10\%.
We select charged pions and kaons that originate from the region
$|\Delta r| < 0.2\,\mathrm{cm}$ and $|\Delta z|< 4\,\mathrm{cm}$ with 
respect to the run dependent 
interaction point, 
 where $\Delta r$, 
$\Delta z$ are the distances of closest approach of $\pi/K$ tracks 
to the interaction vertex in the plane perpendicular to the beam axis and along the
beam axis, respectively. All tracks compatible with the electron
hypothesis ($\sim$ 0.2\% fake rates from pion/kaon) are eliminated.
No attempt has been made to identify muons, which represent a
background of about 2.7\% to the pion tracks.
Candidate $\pi^0$ 
mesons are identified as a pair of isolated ECL clusters with invariant mass in
the window 118 MeV/$c^2$ $<$ $M_{\gamma \gamma}$ $<$ 150 MeV/$c^2$. The energy of each
photon is required to be greater than 30 MeV in the barrel region, defined
as $32^\circ < \theta_\gamma < 128^\circ$, and greater than 50 MeV in the
endcap regions, defined as $17^\circ < \theta_\gamma \leq 32^\circ$ or
$128^\circ < \theta_\gamma \leq 150^\circ$, where $\theta_\gamma$ denotes the 
polar angle of the photon with respect to the direction opposite to the $e^+$ beam.
A mass constrained fit is applied to obtain the 4-momenta of $\pi^0$'s.

Beam gas events are rejected
using the requirements $|P_{z}| < 2$ GeV/$c$ and 0.5 $<$ $ E_{\rm vis}/\sqrt{s}$ $<$ 1.25,
in the $\Upsilon (4S)$ rest frame, where $P_{z}$ and $ E_{\rm vis}$ are the sum of 
the longitudinal momentum and the energy of all reconstructed particles, 
respectively, and $\sqrt{s}$ is the sum of the beam energies in the
$\Upsilon (4S)$ rest frame.

The $\overline{D}^0$ meson is reconstructed through its decay to $K^+ \pi^-$.
A vertex constrained fit is performed 
and the invariant mass is required to
be within 17 MeV/$c^2$ ($\sim$ 3.5 $\sigma$) of the nominal $D$ mass.
$D^{*-}$'s are then reconstructed by combining
the $\overline{D}^0$ with a slow charged pion with $|\Delta r|<$ 2.0\,cm 
and $|\Delta z|<$ 10.0\,cm with respect to the $D$ vertex ($D$ vertex
resolutions are $\sigma_r \sim 0.026$ cm and $\sigma_z \sim$ 0.016 cm). 
The signals of $B \to \overline{D}^{*0}\:(n\pi)$ are 
reconstructed through the decay chain
$\overline{D}^{*0} \to \overline{D}^0 \pi^0$.
Candidate $D^{*-}$ ($\overline{D}^{*0}$) are selected when the reconstructed mass 
difference  between $\overline{D}^{*}$ and $\overline{D}^{0}$ is within 2.0 (2.4) MeV/$c^2$ 
of the nominal mass difference $\Delta M$, which  corresponds to 
$\sim$ 3.5 (3.0)\,$\sigma$ resolution on $\Delta M$. A kinematic 
fit with 
the nominal $\overline{D}^{*}$ mass is applied to obtain 
the 4-momenta of the $\overline{D}^*$ candidate.
 
The $\overline{D}^{*}$ candidate is then combined with $n$ pions to reconstruct the
$B$ meson. All $n$ pions are required to satisfy
$|\Delta r| <$ 0.2 cm and $|\Delta z|<$ 0.8 cm with respect to 
the $\overline{D}$ vertex. 
Continuum ($e^+e^- \to q\overline{q}$, where $q=u,d,s,c$) events are suppressed 
with the criterion,
\mbox{$|\cos\theta_{\rm thrust}| <$ 0.8}, where $ \theta_{\rm thrust}$ is the angle 
between the thrust axis of the $B$ candidate daughters and the thrust 
axis of the 
remaining tracks and isolated ECL clusters. 

\begin{figure}[htbp]
 \begin{center}
  \includegraphics[width=.45\textwidth]{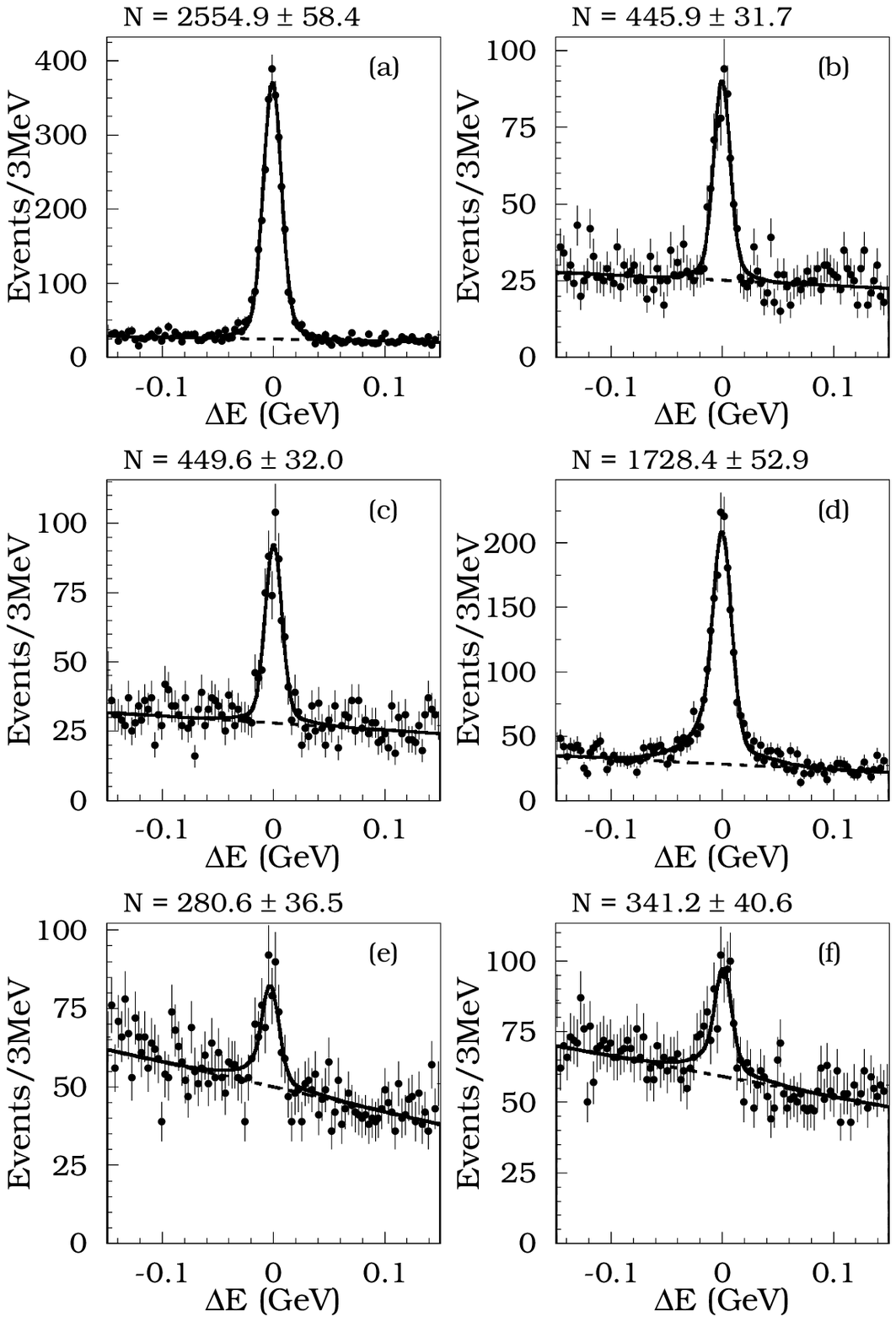}
  \caption{$\Delta E$ distributions for six $\overline{D}^* (n\pi)$ combinations:
          (a) $\bdthrpi$,
          (b) $\bdfourpi$,
          (c) $\bdfivepi$,
          (d) $\bdzethrpi$,
          (e) $\bdzefourpi$ and 
          (f) $\bdzefivepi$. Points with error bars are the observed 
          events in data, solid lines are the results from the fit and dashed
          lines represent the background components.}
  \label{fig:mceff1xall}
 \end{center}
\end{figure}

The signal can then be identified by two kinematic variables calculated in
the $\Upsilon (4S)$ rest frame. The first is the energy difference, 
$\Delta E ~=~E_{D^{*}} + \sum_{i=1}^{ n} E_{ \pi_{i}} - E_{\rm beam}$, 
where $E_{D^{*}}$ is 
the energy of the $\overline{D}^{*}$ candidate, $ E_{ \pi_{i}}$ is the energy of 
the $i$th pion in the $n\pi$ system and 
$E_{\rm beam} =\sqrt{s}/2$ (Fig. \ref{fig:mceff1xall}). 
The second variable is the beam-energy 
constrained mass, $M_{\rm bc}$ = $\sqrt{E^2_{\rm beam} - |\vec{P}_{D^{*}} + 
\sum_{i=1}^{ n} \vec{P}_{\pi_i}|^2}$, where $\vec{P}_{D^{*}}$ 
and $ \vec{P}_{\pi_{i}}$ are momentum vectors of the $\overline{D}^{*}$ candidate, 
and the $i$th pion in the $n\pi$ system.
Typical
resolutions for these variables are 8.5\,MeV and 2.7\,MeV/$c^2$, respectively.
In the extraction of the signal yield,
we require 5.273 GeV/$c^2$ $ <M_{\rm bc} <$ 5.288 GeV/$c^2$ 
and $M_{\rm bc} > $ 5.27 GeV/$c^2$  for  $D^{*-}\:(n\pi)$  and $\overline{D}^{*0}\:(n\pi)$,
respectively,  and we fit the 
$\Delta E$ distribution from $-150$ MeV to $150$ MeV.
 In many $B$ decay analyses, the $\Delta E$ distribution includes peaks
 or other structures due to related $B$ decays with
 an additional particle, one particle less than in
 the mode under study or misidentification of a particle from
 a topologically similar decay mode.
 In this case, we do not observe such structures
 within our fitting range. We do not use the $M_{\rm bc}$ distribution
 to obtain signal yields, because peaking backgrounds in that distribution
 cannot be distinguished from signal.
Selected events contain multiple $B$ candidates with a multiplicity
depending on the signal channels, which varies from 1.2 to 1.8.
For each event we choose a unique $B$ candidate,
taking the combination resulting in the minimum
value of $((M_{\rm bc} - M_{ B})/\sigma_{M_{\rm bc}})^2 + (\Delta M_{ D}/\sigma_{M_{D}})^2 + (\Delta M_{D^*}/\sigma_{ M_{ D^*}})^2$.

We have studied continuum events and other $B$ decays as possible sources of 
background. 
Background due to continuum events is studied by analysing
the 16 fb$^{-1}$ off-resonance data and there we do not find any peak near 
$\Delta E$ = 0. 

$B$ decays (both signal and background) are studied with 
Monte Carlo (MC) event samples. 
MC events are generated using the $QQ$ event generator \cite{qq98} 
with a phase space distribution for the $n\pi$ system, and the response
of the Belle detector is simulated by a GEANT3-based program
\cite{geant3}. The simulated events are then reconstructed and analysed 
with the same procedure as is used for the real data.

We have investigated the possibility of reconstructing
$B \to \overline{D}^{*-(0)} (n' \pi) (m \pi^0)$ channels as
$D^{*-(0)} (n\pi)$ due to the loss and/or addition of pions,
where $n = 3, 4$ or $5$, $n' = n$, $n \pm 1$ and $m = 0$ or $1$
and here we observe a linear background without
any structure around $\Delta E$ = 0.
We have also studied 
$B \to \overline{D}^{*0}$($\to$ $\overline{D}^0 \gamma$)$(n\pi)$ MC
events reconstructed as 
$B \to \overline{D}^{*0}$($\to$ $\overline{D}^0 \pi^0$)$(n\pi)$ and
$D^{*-(0)}\,(n\pi)$ events reconstructed as $D^{*0(-)}\:(n\pi)$.
In each case, we find a $\Delta E$ distribution peaked at zero, but with a
larger width than the corresponding signal distribution.
We also consider backgrounds due to $B \to \overline{D}^0 (n\pi)$ and
$B \to \overline{D}^{0} (n\pi)\pi^0$ with $n \geq 4$. Such decays have not yet been
observed; assuming branching fractions equal to those of the
corresponding $B\to \overline{D}^{*0} (n\pi)$ and $B\to \overline{D}^{*0} ((n+1)\pi)$ modes,
respectively, we find a small background to $D^{*-} (n\pi)$ for 
$n=3,4$, and a negligible contribution to the other final states. 
Finally, we fit MC signal distributions without and with contributions 
from the various feed-across backgrounds. The ratio of those two 
signal yields ($F_r$) depends on the signal 
channels and varies from 0.99 to 0.87. Observed signal yields are corrected 
with the corresponding $F_r$ when extracting branching fractions.

We have studied the following six $B$ decay modes: $\bdthrpi$, $\bdzethrpi$,
$\bdfourpi$, $\bdzefourpi$, $\bdfivepi$ and $\bdzefivepi$. Figure 
\ref{fig:mceff1xall} shows the $\Delta E$ distributions for these decay 
modes. 
Statistically significant structures near $\Delta E$ =0 are observed,
We have also checked the corresponding $\Delta E$
distributions for events in the $M_{\rm bc}$ sideband region (5.23 GeV/$c^2$ 
$<M_{\rm bc}<$ 5.26 GeV/$c^2$): no structure is observed.
For all the decay modes under study, backgrounds are fitted with a linear function.
The signal shape is modelled with a sum of two Gaussian distributions
for $B \to \overline{D}^{*-} (n\pi)$. The $\overline{D}^{*0} (n\pi)$ signal is fitted 
with a sum of the Crystal Ball lineshape (CB) \cite{tomasz},
\begin{eqnarray*}
F(\Delta E)&=&A \cdot\exp\left({-\,0.5 \cdot \left( \frac{\Delta E}
 {\sigma_{\Delta E}}\right)^{2}}\right) \\
& & \hspace*{8mm} {\rm for}~ \Delta E 
 ~\geq -\,\alpha \sigma_{\Delta E} \\
            &=&A \cdot\exp\left(-\,0.5\cdot \alpha^{2}\right) \cdot
 \left[1-\frac{\alpha}{n}\cdot\frac{\Delta E}{\sigma_{\Delta E}} -
 \frac{\alpha^{2}}{n}\right]^{-n} \\
 & &  \hspace*{8mm} {\rm for}~ \Delta E ~< -\,\alpha \sigma_{\Delta E}
\end{eqnarray*}
and a 
Gaussian. The $n$ and $\alpha$ parameters of the CB and 
the fractional area of the second Gaussian are fixed from fits to 
the MC sample.
Signal yields obtained from the fit are
summarised in Table \ref{tab:branch}. 

\begin{table}[htbp] 
\begin{center}
\caption{Measured signal yields and branching fractions.}
\vspace{1mm}
\begin{ruledtabular}
\begin{tabular}{|l|rrr|rrrrr|}
\multicolumn{1}{|c|}{Channel} & 
\multicolumn{3}{c|}{Signal yield } &
\multicolumn{5}{c|}{$\br \times 10^3$} \\ \hline

$\bdthrpi    $ & 2554.9 &$\pm$& 58.4 &6.81  &$\pm$&0.23 &$\pm$&0.72 \\ 
$\bdfourpi   $ &  445.9 &$\pm$& 31.7 &2.56  &$\pm$&0.26 &$\pm$&0.33 \\
$\bdfivepi   $ &  449.6 &$\pm$& 32.0 &4.72  &$\pm$&0.59 &$\pm$&0.71 \\
$\bdzethrpi  $ & 1728.4 &$\pm$& 52.9 &10.55 &$\pm$&0.47 &$\pm$&1.29 \\
$\bdzefourpi $ &  280.6 &$\pm$& 36.5 &2.60  &$\pm$&0.47 &$\pm$&0.37 \\
$\bdzefivepi $ &  341.2 &$\pm$& 40.6 &5.67  &$\pm$&0.91 &$\pm$&0.85 \\ 
\end{tabular} 
\end{ruledtabular}

\label{tab:branch}
\end{center}
\end{table}

The signal efficiency depends on the invariant mass ($M_{n\pi}$) of 
the $n\pi$ system. Signal efficiencies are calculated 
from MC event samples in 100 MeV/$c^2$ bins of $M_{n\pi}$. 

\begin{figure}[htbp]
 \begin{center}
  \includegraphics[width=.48\textwidth]{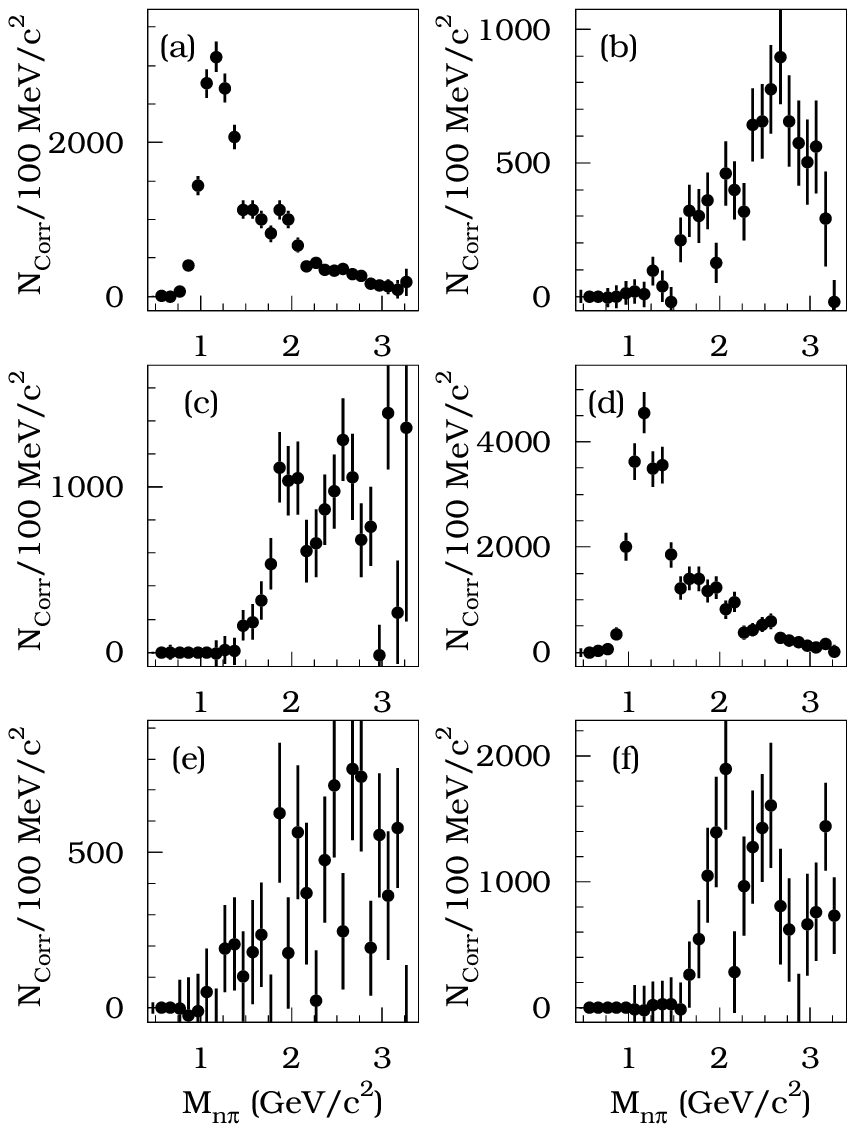}
  \caption{Efficiency corrected $n\pi$ invariant mass spectra for 
          (a) $\bdthrpi$,
          (b) $\bdfourpi$,
          (c) $\bdfivepi$,
          (d) $\bdzethrpi$,
          (e) $\bdzefourpi$ and 
          (f) $\bdzefivepi$.}
  \label{fig:effcorrnum}
 \end{center}
\end{figure}

In order to obtain the signal yield in each $M_{n\pi}$ bin,
we perform a fit to $\Delta E$ distributions in each bin to
avoid the possible uncertainty due to different
background shapes in signal and sideband regions of $M_{\rm bc}$.
Efficiency corrected $M_{n\pi}$ spectra are shown in 
Fig. \ref{fig:effcorrnum}.
Invariant mass distributions of the $3\pi$ system ($M_{(3\pi)}$) in 
$B \to \overline{D}^* (3\pi)$ decays show clear evidence for the
presence of $a_1$ (Fig. \ref{fig:effcorrnum}(a,d)).
The $M_{4\pi}$ distributions in Fig. \ref{fig:effcorrnum}(b,e) do
not show any resonant structure. $M_{5\pi}$ distributions in 
Fig. \ref{fig:effcorrnum}(c,f) show a peak around 2 GeV/$c^2$ which,
however, does not correspond to any known resonance. 
We also search for narrow resonances in the $M_{5\pi}$
distributions, after the subtraction of $\Delta E$
sidebands. Apart from a peak due to the $D_s^+$, no narrow
structure is observed.

Inclusive branching fractions obtained for the six modes are listed  
in Table \ref{tab:branch}, where the first error is statistical and
the second error is systematic. This is the first measurement of the 
branching 
fractions of $\bdfourpi,\:\bdfivepi$ and $\bdzefivepi$. The branching
fractions for the decay modes $\bdthrpi$, $\bdzethrpi$ and $\bdzefourpi$ are
measured with better precision than in previous 
studies \cite{pdg2002}. We do not exclude contributions from exclusive 
channels $B \to \overline{D}^{*} D_s^+$, $B \to \overline{D}^{*} D^{*+}$, 
$B \to \overline{D}^{*} D^+$, $B \to \overline{D}^{*} D^{*+} K_S$ etc. The largest 
contributions of this type to inclusive $\overline{D}^* (3\pi)$ and 
$\overline{D}^* (5\pi)$
are expected to come from $\overline{D}^{*} D_s^+$, with 
$D_s^+ \to (3\pi)^+$ (${\cal{B}} = 1.01 \pm 0.28$\% \cite{pdg2002}) and 
$D_s^+ \to (5\pi)^+$ (${\cal{B}} = 0.65 \pm 0.18$\% \cite{pdg2002}), respectively.
We conduct a search for these modes by fitting invariant mass distributions 
of $n\pi$, $M_{n\pi}$ for signal events ($|\Delta E| <$ 30 MeV)
 and find yields of 27.8$\pm$7.7,
15.6$\pm$5.4, 8.4$\pm$3.9 and 11.3$\pm$5.4 in  $\bdthrpi$, $\bdzethrpi$, 
$\bdfivepi$ and $\bdzefivepi$ channels, respectively. 
Within the statistical errors, these results are consistent with PDG 
expectations. The expected 
contribution from other modes are smaller than $B \to \overline{D}^{*} D_s^+$ by a
factor $\sim$ 4. 

\begin{table}[htbp] 
 \begin{center}
\caption{Contributions of systematic uncertainties (in \%).}
\vspace{1mm}
\begin{ruledtabular}
  \begin{tabular}{|l|rcc|ccc|}
& \multicolumn{3}{c|}{$D^{*-}$} & \multicolumn{3}{c|}{$\overline{D}^{*0}$} \\ \cline{2-7}
& $3\pi $ & $4\pi $ & $5\pi$ & $3\pi $ & $4\pi $ & $5\pi$  \\ \hline
Track finding& 8.7 &11.0 &12.7 & 5.4 & 6.6 & 7.9 \\
Slow $\pi^0$ & -& -& -&7.0 & 7.0  & 7.0 \\
PID         &  4.9 & 5.0 & 4.8 & 5.0 & 5.2 & 5.3 \\
Branchings &    2.5 & 2.5 & 2.5 & 5.3 & 5.3 & 5.3 \\
Feed-across &  0.8 & 1.0 & 2.0 & 0.9 & 1.4 & 2.2 \\
Fitting &      0.5 & 1.2 & 1.4 & 2.4 & 4.0 & 5.5 \\
Selection &    2.5 & 3.7 & 4.9 & 3.3 & 5.7 & 4.7 \\ 
MC            &0.8 & 0.8 & 2.2 & 1.2 & 1.1 & 1.7 \\ 
$N_{B\overline{B}}$&0.5 & 0.5 & 0.5 & 0.5 & 0.5 & 0.5 \\ \hline
Total         &10.6 &13.0  & 15.0 &12.3  & 14.1 & 15.0 \\ 
\end{tabular}
\end{ruledtabular}
\label{tab:system}
\end{center}
\end{table}

The systematic uncertainty is obtained from a quadratic sum of
        nine terms, which are shown in Table \ref{tab:system}: the 
        uncertainty in (a) the track
        finding efficiency, ranging from 1\% for high momentum tracks 
        to 8\%
        for 80 MeV/$c$ pions, calculated from partially reconstructed 
        $D^{*-} \to \overline{D}^0 (\to K^0_s (\to \pi^+ \pi^-) \pi^+ \pi^-) \pi^-$
        events and a track
        embedding study; (b) the slow $\pi^0$ finding efficiency; (c) $K/\pi$
        selection efficiencies (PID), calculated using 
        $D^{*-} \to \overline{D}^0 (\to K^+ \pi^-) \pi^-$
        events; (d) branching fractions of 
        $D^{*-}(\overline{D}^{*0}) \to \overline{D}^0 \pi^{-(0)}$, 
        $\overline{D}^0 \to K^+ \pi^-$ and $\pi^0 \to \gamma \gamma$; 
        (e) the uncertainty in the feed-across from other decay modes
	which is calculated by changing the relative branching 
        fractions of backgrounds to signal by one sigma of their errors; 
        (f) the uncertainties due to the choice of signal shape and 
        histogram  binning. $\Delta E$ distributions are also fitted with
        $(i)$ single Gaussian (CB), 
        $(ii)$ modified Gaussian (CB), where Gaussian power has changed from 2 to 1.4, 
        $(iii)$ asymmetric Gaussian,
        $(iv)$ modified Gaussian (CB) + Gaussian and
        $(v)$ asymmetric Gaussian + Gaussian for
        $D^{*-}\:(n\pi)$ ($\overline{D}^{*0}\:(n\pi)$) signals.
	Ratio of fitted yields in data and MC signals are calculated for
        these functions and the variation of these ratio is taken as 
        the systematic error due to fitting function; 
        (g) the uncertainties due to selection criteria, which
        are calculated by varying those criteria by one sigma of their errors,
        (h) limited MC statistics 
        and (i) the uncertainty in the total
        number of ${B\overline{B}}$ events ($N_{B\overline{B}}$).

In summary, we have made first observations of decay channels 
$\bdfivepi$, $\bdfourpi$ and $\bdzefivepi$ using 152 million 
$ B \overline{B}$ events.
We measure inclusive branching fractions for these three decay modes.
We have also made precise measurements of the branching fractions for the
decay channels $\bdthrpi, \: \bdzethrpi$ and $\bdzefourpi$.

We thank the KEKB group for the excellent operation of the
accelerator, the KEK Cryogenics group for the efficient
operation of the solenoid, and the KEK computer group and
the NII for valuable computing and Super-SINET network
support.  We acknowledge support from MEXT and JSPS (Japan);
ARC and DEST (Australia); NSFC (contract No.~10175071,
China); DST (India); the BK21 program of MOEHRD and the CHEP
SRC program of KOSEF (Korea); KBN (contract No.~2P03B 01324,
Poland); MIST (Russia); MESS (Slovenia); Swiss NSF; NSC and MOE
(Taiwan); and DOE (USA).


\begin{thebibliography}{99}

\bibitem{pdg2002}  S. Eidelman {\it et al.} (Particle Data Group), 
                   Phys. Lett. B{\bf 592}, 1 (2004).
\bibitem{cleochg}  CLEO Collaboration, G. Brandenburg {\it et al.}, 
                   Phys. Rev. D{\bf 61}, 072002 (2002).
\bibitem{cleorhoa1}ARGUS Collaboration, H. Albrecht {\it et al.},  
                   Z. Physik C{\bf 48}, 543 (1990); 
                   CLEO Collaboration, M. S. Alam {\it et al.}, 
                   Phys. Rev. D{\bf 50}, 43 (1994);
                   CLEO Collaboration, K.W. Edwards {\it et al.}, 
                   Phys. Rev. D{\bf 65}, 012002 (2002).
\bibitem{llw}      Z. Ligeti, M. Luke and M. B. Wise, Phys. Lett. B{\bf 507}, 
                   142 (2001).
\bibitem{belledet} Belle Collaboration, A. Abashian {\it et al.}, 
                   Nucl. Instrum. Methods Phys. Res. A{\bf 479}, 117 (2002).
\bibitem{kekb}     S. Kurokawa and E. Kikutani, 
                   Nucl. Instrum. Methods Phys. Res. A{\bf 499}, 1 (2003).
\bibitem{qq98}     The $QQ$ $B$ meson decay event generator was developed by 
                   the CLEO collaboration, 
                   http://www.lns.cornell.edu/public/CLEO/soft/QQ
\bibitem{geant3}   CERN Program Library Long Writeup, W5013, CERN, 1993.
\bibitem{tomasz}   T. Skwarnicki, Ph. D. thesis, Institute for Nuclear Physics,
                   Krakow, 1986.
\end{thebibliography}
\end{document}